\documentclass[prb,reprint,twocolumn,showpacs,superscriptaddress,floatfix]{revtex4-1}
\usepackage{graphicx,amssymb,amsmath,bbold,bm,wasysym,psfrag,color,tikz}
\graphicspath{{img/}}
\usepackage{mathtools}
\usepackage[normalem]{ulem}

\begin{document}
	
\title{Polarization Plateaus in Hexagonal Water Ice $I_h$}
\author{Matthias Gohlke}
\affiliation{Max-Planck-Institut f\"ur Physik komplexer Systeme, 01187 Dresden, Germany}
\author{Roderich Moessner}
\affiliation{Max-Planck-Institut f\"ur Physik komplexer Systeme, 01187 Dresden, Germany}
\author{Frank Pollmann}
\affiliation{Technische Universit\"at M\"unchen, 85747 Garching, Germany}
\affiliation{Munich Center for Quantum Science and Technology (MCQST), 80799 M\"unchen, Germany}
\date{\today}     
	
\begin{abstract}
    The protons in water ice are subject to so called \emph{ice rules} resulting in an extensive ground state degeneracy.
    We study how an external electric field reduces this ground state degeneracy
    in hexagonal water ice I$_h$ within a minimal model.
    We observe polarization plateaus when the field is aligned along the $[001]$ and $[010]$ directions.
    In each case, one plateau occurs at intermediate polarization with reduced
    but still extensive degeneracy.
    The remaining ground states can be mapped to dimer models on the honeycomb
    and the square lattice, respectively.
    Upon tilting the external field, we observe an order-disorder transition of Kasteleyn type
    into a plateau at saturated polarization and vanishing entropy.
    This transition is investigated analytically using the Kasteleyn matrix
    and numerically using a modified directed-loop Monte Carlo simulation.
    The protons in both cases exhibit algebraically decaying correlations.
    Moreover, the features of the static structure factor are discussed.
\end{abstract}
	\maketitle

\section{Introduction} %

\begin{figure}
    \includegraphics[width=0.98\linewidth]{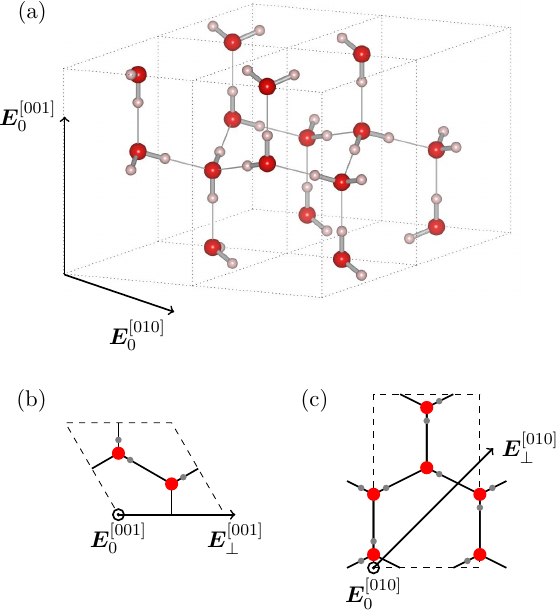}
	\caption{
        (a) Illustration of the structure of hexagonal ice I$_h$.
        An external field along $\bm E^{[001]}_0 \parallel [001]$
        or $\bm E^{[010]}_0 \parallel [010]$ is applied with an additional tilt in the directions:
        (b) $\bm E^{[001]}_\perp \parallel [100]$
        and (c) $\bm E^{[010]}_\perp \parallel [101]$.
        Oxygen ions are drawn in red, protons in gray.
    }
	\label{fig:fieldandstruct}
\end{figure}

Water exists in a huge variety of different phases including liquid, gas, as well as numerous crystalline and amorphous phases.
The most common crystalline phase of water on earth is the hexagonal phase \emph{ice I$_h$}, cf. Fig.\ref{fig:fieldandstruct}(a).
Bernal and Fowler proposed that ice I$_h$ is a molecular solid formed by H$_2$O molecules\cite{bernal_theory_1933}
which obey the \emph{ice rules}:
a single proton is placed on each of the four oxygen-oxygen link per oxygen site.
There are two protons bonded covalently to each oxygen ion forming H$_2$O molecules
and the two other protons form hydrogen bonds.
As such the ice rules impose local constraints
that can be expressed as a zero divergence condition on a lattice
and give rise to emergent gauge fields.\cite{huse_coulomb_2003,hermele_pyrochlore_2004}
These local constraints prevent the H$_2$O molecules from ordering
and leave a massively degenerate proton subsystem with a finite residual entropy.%
\cite{pauling_structure_1935,giauque_entropy_1936}
Consequently, ice I$_h$ is considered to be a prototype of geometrical frustration.

Because of the third law of thermodynamics, the equilibrium residual entropy is expected to vanish as temperature approaches zero,
however, the nature of the true ground state of ice I$_h$ 
is not yet settled.\cite{minagawa_ferroelectric_1981, minagawa_phase_1990, abe_note_1993,kirov_proton_1996,jackson_thermallystimulated_1997,iedema_ferroelectricity_1998, lekner_energetics_1998, rick_simulations_2005, parkkinen_ice_2014} 
On one hand,
a transition with a significant reduction in entropy is observed in KOH doped ice I$_h$ at $T=72K$.
\cite{kawada_dielectric_1972,tajima_phase_1982, matsuo_calorimetric_1986,suga_facet_1997}
On the other hand relaxation times grow beyond reasonable observation times
for pure ice below $T\approx 110K$,\cite{wooldridge_proton_1988,suga_facet_1997}
obscuring any thermodynamic transition at lower $T$.
Yet, experimental signatures of correlated proton hopping processes were reported\cite{bove_anomalous_2009,yen_dielectric_2015} 
enabling the protons to possibly form a quantum proton liquid,\cite{castroneto_ice_2006,benton_classical_2016}
which would be a strong analogy to quantum spin liquids.\cite{anderson_resonating_1973}

A magnetic equivalent of the ice rules exists in \emph{spin ice} materials.%
\cite{harris_geometrical_1997,ramirez_zeropoint_1999,bramwell_spin_2001} 
Due to strong crystal fields, local magnetic moments are aligned parallel or antiparallel to links
in and out of the tetrahedrons of a pyrochlore lattice.
The local moments have to satisfy a modified ice rule: two spins point into a tetrahedron and two point out.
Analog to ice,
the constraint is equivalent to the zero divergence condition of classical electrodynamics,
Gauss' law, on a lattice.

In both cases, water ice or spin ice, the local constraints 
manifest themselves in certain features of scattering experiments.
The ice rules cause an angle dependence of the scattering intensity around lattice Bragg points
resulting in a bow-tie-like structure.%
\cite{youngblood_correlations_1980,li_diffuse_1994}
Such scattering maps contain information on proton-proton correlations,
which have been computed for ice I$_h$ using a large-$N$ theory\cite{isakov_analytical_2015}
or a projection operator approach.\cite{benton_classical_2016}

In this article, we study the reduction of the residual entropy by applying an external electric field
on the extensively degenerate ground state of the proton configurations. 
Generically, the degeneracy is expected to decrease upon applying an external field,
but details depend on the direction of the field and which lattice symmetries the field breaks.
The degeneracy remains extensive in certain field directions,
e.g., when the field confines the remaining degrees of freedom to a stack of decoupled planar layers
such as for pyrochlore spin ice in a $[111]$ magnetic field.%
\cite{matsuhira_new_2002,moessner_theory_2003}
A small tilt of the field results in a transition into a long range ordered state.%
\cite{moessner_theory_2003}.

We consider an external electric field that couples to the proton subsystem via a Stark term.
The proton configurations are described by a minimal model of ice I$_h$ satisfying the ice-rules:
a proton can only take one of two positions on a link between two neighboring oxygen atoms
and only a single proton per link exists.
This allows to describe the proton configurations as Ising-like spins aligned parallel to the links.

We focus on fields along two easy axes causing a partial ordering of the protons.
The remaining degrees of freedom reside on decoupled layers mappable to hardcore dimer models
either on an honeycomb or a square lattice, respectively. 
The former is known from pyrochlore spin-ice,\cite{moessner_theory_2003}
whereas the latter has no analog in spin-ice and, thus, is the main focus of this work.
The field effectively reduces the dimensionality of the system
to uncorrelated two-dimensional layers.
While the ground state degeneracy is reduced, it remains extensive.
%
A slight tilt of the field lifts the remaining degeneracy within each layer
and leads to an order-disorder transition first described by Kasteleyn.\cite{kasteleyn_dimer_1963} 
Using the Kasteleyn-matrix approach\cite{kasteleyn_dimer_1963,cohn_variational_2001,kenyon_dimers_2006,kenyon_lectures_2009}
for the emergent planar dimer models,
we investigate the critical behaviour in the thermodynamic limit
and compare to numerical simulations using
a modified directed-worm Monte-Carlo simulation on the three-dimensional model.
Moreover, the static structure factor is obtained from the Monte-Carlo simulation.
It shows typical bow tie or pinch point features,\cite{lacroix_introduction_2011}
that get supplemented by logarithmically diverging peaks within the emergent dimer regime.

The article is structured as follows:
In section~\ref{scn:mod}, the structure of ice I$_h$ and the model are introduced.
An explanation of the modified directed-worm Monte-Carlo method follows in Sec.~\ref{scn:met}.
Results based on thermodynamic observables are presented in Sec.~\ref{scn:pol} 
following a discussion of the emergent dimers on a hexagonal lattice in Sec.~\ref{scn:ice_hex_dim}
and the square lattice in Sec.~\ref{scn:ice_sq_dim}.
The latter is extended by utilizing the Kasteleyn-matrix to derive the dimer densities, cf. Sec.~\ref{scn:ice_dimer}.
Section~\ref{scn:ssf} contains correlations and diffuse scattering maps.

\section{Model} \label{scn:mod}

Ice I$_h$ has a bipartite hexagonal lattice with four oxygen atoms
and eight protons per primitive cell (Fig. \ref{fig:fieldandstruct}).
The oxygen atoms form a four-fold coordinated tetrahedral structure of class \textit{P6$_3$/mmc}.
The following symmetries exist: a sixfold rotation axis, a twofold rotation axis perpendicular to the z-axis
and a horizontal reflection plane. 
The structure can be considered as a stacking of planes in an ABAB pattern,
where B is the reflection of A with respect to a horizontal plane.

Within the unit-cell, eight distinct oxygen-oxygen links exist. 
Each proton has two possible positions arranged symmetrically around the midpoint of a link.
We neglect the displacement of the protons away from the link.
If we assume a single proton per link, which is one of the ice rules,\cite{bernal_theory_1933}
the proton configuration can be represented as a configuration of electric dipoles $\bm d_i$
\begin{equation}
    \bm d_i = s_i d \bm{\hat e_i}~.
\end{equation}
The dipoles can be represented as effective Ising spins $s_i$ with respect to a local z-axis
given by the direction $\bm{\hat e_i}$ of the oxygen-oxygen links.

We consider the following minimal model in terms of electric dipoles or Ising spins, respectively:
\begin{equation}
    H = J \sum_\alpha L_\alpha^2 - \sum_i s_i d \bm E \cdot \bm {\hat e_i}~,
\end{equation}
where $L_\alpha = \sum_{i \in \boxtimes_\alpha} s_i$
is the sum over effective Ising spins on a tetrahedron $\boxtimes_\alpha$.
In choosing the limit $J \rightarrow \infty$,
we restrict the dipole configuration to the ice-rule manifold. 
The second term represents the Stark-coupling of the moment to an external electric field, $\bm E$.
The direction of the unit vectors $\bm{\hat e_i}$ along the links is chosen
such that a positive $s_i$ points from a site on sublattice A to a site on sublattice B. 
In the remainder, the constant $d$ is set to unity.

We apply the field along a) the c-axis $[001]$ and b) $[010]$ with respect to the orthorhombic unit cell,
see Fig.~\ref{fig:fieldandstruct} for an illustration. 
These field directions still leave an extensively degenerate ground state
with finite entropy density. 
In the following, we investigate the case where the field is slightly tilted towards
a) [100] and b) [101],
such that a single lowest energy state exist. 
For the remainder of this article, the notation $\bm E = \bm E_0 + \bm E_\perp$ is used for the external field,
where $\bm E_0$ is the component along the easy axis and $\bm E_\perp$ denotes the tilt component with $\bm E_\perp \ll \bm E_0$.

\section{Modified Directed-Worm Monte Carlo Sampling} \label{scn:met} 

The spin- or proton system is simulated numerically
using a directed loop Monte-Carlo method.\cite{barkema_monte_1998} 
Since a single spin flip would violate the second ice rule,
loops of consecutive spins are created and flipped instead.
In doing so, the simulation remains in the defect-free manifold.
In order to keep the efficiency high, the sampling enters already at the loop creation: 
two defects are created by choosing a single spin randomly with uniform probability and flipping it. 
One of the defects executes a weighted random walk
with weights depending on the energetics of the original local configuration and the possible new local configurations. 
As soon as that defect meets the other defect, the loop is closed and all corresponding spins will be flipped. 
This is also known as the \emph{long loop} algorithm,
which has the property of spanning on average a fraction of the whole system.\cite{jaubert_analysis_2011}
A loop which winds once or multiple times around the system connects different winding sectors.

In order to satisfy detailed balance, backtracking of a defect must also be taken into account. 
If the defect tracks all the way back to its initial creation,
no update occurs and a new initial spin randomly chosen.

In most cases,
only the change of the local configuration on a single tetrahedron is taken into account
when computing the probabilities to move the defect on site. 
However, this choice makes the algorithm highly inefficient for the field along $[010]$. 
Consider two neighboring vertices connected by a spin perpendicular to the field: 
the defect steps forward at the first vertex with high probability as the change in configuration leads to a gain in energy.
At the next vertex, the defect has only a small probability to move forward as it would cost nearly the same energy
and moving back is more likely.
Thus, with decreasing temperature the probability of oscillating between the two vertices increases 
until the probability converges to one. 
This issue can be circumvented by considering double vertices of two neighboring tetrahedra
sharing this particular spin perpendicular to the field. 
By doing so, the energy gain and cost are taken into account within one step.

The weighted random walk of the defect will be performed based on local heat bath sampling. 
Moving a defect through the smallest unit, either a single tetrahedron or a double tetrahedron,
will change its local configuration from $i$ to $j$.
The probability is given by
\begin{equation}
	p(i \rightarrow j) = \frac{z(i \rightarrow j)}{\sum_k z(i \rightarrow k)}~,\nonumber
\end{equation}
where $z(i \rightarrow j) =  e^{-\beta(\epsilon_j - \epsilon_i)}$ is the Boltzmann weight of a local configurational change.
The sum in the denominator runs over all possible configurations $k$
which can be achieved by forward steps and changing the local configuration ($i \ne k$) 
or by backtracking and retaining the local configuration ($i=k$).
The energy $\epsilon_i$ of a local configuration is either given by
\begin{equation}
    \epsilon_\alpha = - \frac{1}{2} \bm E \sum_{i \in \boxtimes_\alpha} s_i \bm{\hat e_i}, \nonumber
\end{equation}
where the sum runs over all links of a single vertex $\boxtimes_\alpha$,
or, for the modified algorithm, by 
\begin{equation}
    \epsilon_\alpha = - s_{\alpha,\text{int}} \bm E \cdot \bm{\hat e_{\alpha,\text{int}}}
                    - \frac{1}{2} \sum_{\text{ext}~i \in D_\alpha} s_i \bm E \cdot \bm{\hat e_i}~, 
    \label{eqn:energy_doublevertex}
\end{equation}
summing the Stark energy contribution from the internal spin of a double vertex $D_\alpha$. 
The (external) spins are weighted by a factor $1/2$, since they are shared by two neighboring (double) tetrahedron. 

The protocol for the simulation is as follows: 
An initial configuration is chosen that obeys the ice rules everywhere,
loop creations and spin flips at infinite temperature follow
in order to randomize the initial state. 
After then equilibrating at a temperature of choice,
the auto-correlation time is measured and determines how many loop iterations will be done 
between each measurement at the particular temperature.
This procedure is repeated for all temperatures of interest. 
The field is held constant over the whole procedure. 
Simulations are done on systems with up to $160 \times 160 \times 80$ hexagonal unit cells and with periodic boundary conditions.
 
\section{Polarization Plateaus} \label{scn:pol} 
    
\begin{figure}[tb]
    \includegraphics[width=\linewidth]{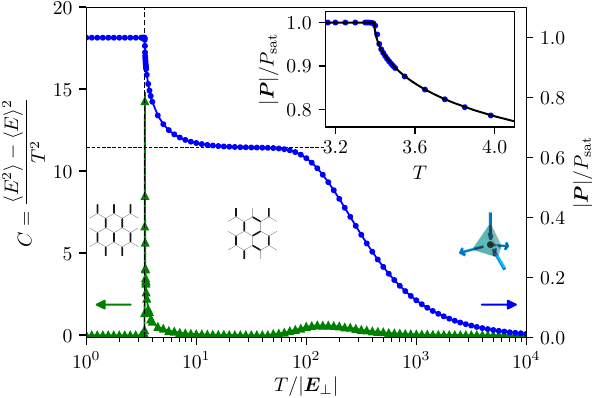} 
    \caption{
        Relative polarization (blue dots) and specific heat (green triangles)
        versus temperature for a field applied along $\bm E = (1,0,100)$.
        Three regimes exist:
        I) a high-$T$ regime with only small polarization only subject to the ice rules (tetrahedron with a two-in-two-out configuration),
        II) an intermediate polarization plateau (random dimers),
        and III) a low-$T$ regime with fully saturated polarization and long-range order (ordered dimers).
        The polarization plateau has a relative polarization 
        of $|\bm P|/P_\text{sat} = \frac{2}{\sqrt{10}}\approx 0.63$ (horizontal dashed line).
        The degrees of freedom within the plateau form a stack of two-dimensional planes,
        where for each of the planes a microscopic description
        in terms of dimers on a honeycomb lattice exist.
        A Kasteleyn transition into a fully ordered state
        occurs at $T_K=3.96$ (vertical dashed line).
        The inset shows the relative polarization near the Kasteleyn transition.
        The Kasteleyn matrix approach enables us
        to compute the relative polarization (solid black line).
        Excellent agreement is found with the numerical data (blue dots).
    }
    \label{fig:ice_P_C_hex}
\end{figure}

\begin{figure}[tb]
    \includegraphics[width=\linewidth]{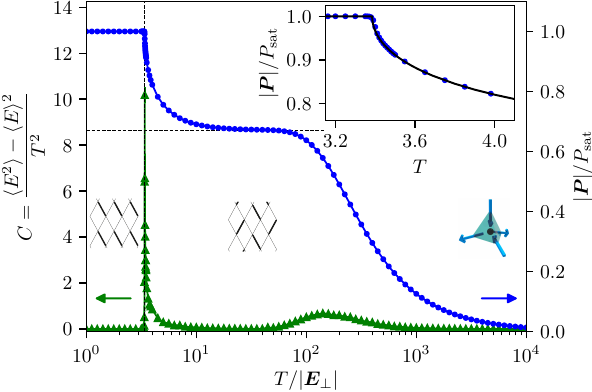} 
    \caption{
        Same plot as in Fig.~\ref{fig:ice_P_C_hex} but for a field applied along 
        $\bm E = \left(\frac{1}{\sqrt{2}}, 100, \frac{1}{\sqrt{2}}\right) \approx (0.71, 100, 0.71)$.
        The Kasteleyn transition occurs at $T_K=3.385$ (vertical dashed line).
        The polarization in the intermediate plateau points along $[010]$
        with magnitude $|\bm P|/P_\text{sat} = \frac{3}{2\sqrt{5}}\approx 0.67$
        (horizontal dashed line). 
        Here, the degrees of freedom within the plateau form a stack of two-dimensional planes,
        where for each of the planes a microscopic description
        in terms of dimers on a square lattice exists.
        The inset shows the relative polarization near the Kasteleyn transition of the effective dimers into a fully ordered state.
        The numerical data agrees with the polarization obtained analytically (solid black line)
        utilizing the Kasteleyn-matrix approach, cf. Sec.~\ref{scn:ice_dimer}.
    }
    \label{fig:ice_P_C_sq}
\end{figure}

The numerically obtained polarization and the specific heat are shown in
Fig.~\ref{fig:ice_P_C_hex} for (i) $\bm E_0$ along $[001]$ with a tilt $\bm E_\perp$ along $[100]$,
and in Fig.~\ref{fig:ice_P_C_sq} for (ii) $\bm E_0$ along $[010]$ with a tilt $\bm E_\perp$ along $[101]$.
The applied fields are
\begin{align}
    \text{(i)} \quad &\bm E = (1,0,100) \nonumber \\
    \text{(ii)} \quad &\bm E = \left(\frac{1}{\sqrt{2}}, 100, \frac{1}{\sqrt{2}}\right)
                            \approx (0.71, 100, 0.71)~. \nonumber 
\end{align}
$|\bm E_\perp|$ is set to be two orders of magnitude smaller than $|\bm E_0|$.

As a consequence of the two energy scales set by $\bm E_0$ and the tilt $\bm E_\perp$,
three temperature regimes can be distinguished. 
For temperatures $T \gg |\bm E_0|$ the physics is solely determined by the ice rules
and, consequently, the polarization is close to zero. 
Upon cooling down to $|\bm E_\perp| \ll T < |\bm E_0|$, a crossover into an intermediate polarization plateau is observed.
In the intermediate plateau, the spin sublattice with the maximal projection along the field gets pinned.
The other spins remain free and form decoupled, two-dimensional layers.
Thus, a dimensional reduction of the physics from three to two dimensions occurs due to the external field.
The polarization is determined by the fixed spins and is illustrated as horizontal dashed lines in Fig.~\ref{fig:ice_P_C_hex}
and \ref{fig:ice_P_C_sq}.
%


For both field directions,
the crossover between high-temperature and intermediate plateau
is in close relation to spin ice.\cite{isakov_magnetization_2004} 
Starting from the intermediate plateau,
the polarization can only be reduced by introducing string-defects.
Such string-defects are composed of concurrent spins and span the entire system in order to not violate the ice rules.
Spins along a string-defect are flipped once a temperature is reached
at which the gain in entropy compensates for the increase in internal energy induced by the defect.
The two-dimensional layers, which are decoupled at low temperatures,
begin to interact at this crossover temperature and three-dimensionality gets restored.
Further increase of the temperature results in multiple string-defects that exhibit 
an interaction between string-defects of entropic origin.

%
%


A second polarization plateau with saturated polarization is approached at temperatures below the intermediate regime.
Near $T_K \approx 3.4$ the specific heat diverges for $T>T_K$,
indicating a continuous phase transition. 
In contrast, the specific heat vanishes for $T<T_K$, which would indicate a first order phase transition. 
In fact, any fluctuations are suppressed below the transition as they are only caused by string defects
spanning the entire system.
Such a so called Kasteleyn transition is well known in the context of dimer models.\cite{kasteleyn_dimer_1963}

We continue with a more detailed discussion of the intermediate plateau
and the Kasteleyn-type transition into the fully polarized phase for the two field directions.
A one-to-one mapping from spin configurations within the decoupled layers to hardcore dimers exists.
The discussion of the emergent dimer models
is given separately for the two field directions studied.
The first field direction (i) leads to decoupled layers of dimers on a hexagonal lattice,
and the second field direction (ii) gives rise to dimers on a square lattice. 

\subsection{Intermediate plateau for $\bm E \parallel [001]$: Dimer model on a hexagonal lattice} \label{scn:ice_hex_dim}

\begin{figure}[tb]
    \includegraphics{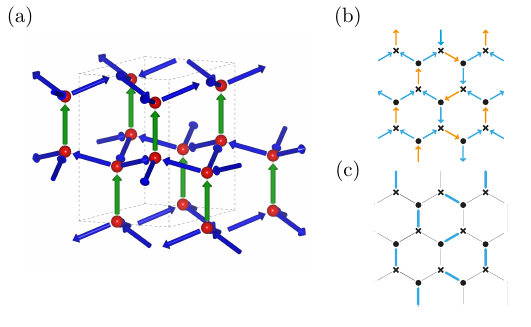}
    \caption{
        Upon applying a field along $[001]$
        and lowering the temperature below the corresponding energy scale,
        the spins with the largest projection onto the field are pinned.
        These are highlighted as green spins in (a).
        The remaining degrees of freedom, blue spins in (a), form a bipartite honeycomb lattice
        with two-in-one-out or one-in-two-out configurations, see (b).
        (c) shows a one-to-one map between the low energy manifold of spin configurations
        and dimers on the edges of a honeycomb lattice.
    }
    \label{fig:ice_spins001}
\end{figure}

Applying an electric field along $[001]$ pins the spins parallel to the field, the green spins in Fig. \ref{fig:ice_spins001}(a),
as soon as the ratio $|\bm E_0|/T$ is sufficiently large.
The remaining degrees of freedom, the blue spins,
reside on the edges of two dimensional honeycomb lattices with alternating two-in-one-out or one-in-two-out vertices%
\footnote{Such a partial ordered state of ice I$_h$ was proposed early on in a different context.\cite{rundle_structure_1953}}
We can map the blue spins to dimers on the same honeycomb lattice
by assigning a dimer to each minority spin,
which are highlighted in orange in Fig. \ref{fig:ice_spins001}(b).
The hexagonal layers form an 'ABAB'
stacking with inverted two-in-one-out or one-in-two-out vertices. 
Apart from the stacking,
this intermediate state is identical to the well studied kagome ice state of pyrochlore spin ice
in a $[111]$ magnetic field.\cite{moessner_theory_2003,isakov_magnetization_2004}

Pinning one out of four spin per site reduces the entropy,
however, the number of dimer coverings of the hexagonal lattice still grows exponentially
and results in a residual entropy.
In case of equally weighted dimers,
the value of the entropy is reduced to
$S/k_B \approx 0.1615$ per site\cite{wu_remarks_1968}
from the total residual entropy of I$_h$, $S/k_B \approx 0.410$ per site.%
\cite{nagle_lattice_1966,isakov_magnetization_2004,vanderstraeten_residual_2018}

Closely packed configurations of dimers on the edges of a honeycomb lattice
exhibit a transition between a long-range ordered phase
and a disordered phase,\cite{kasteleyn_dimer_1963,wu_remarks_1968}
The disordered phase exhibits an algebraic decay of two-point correlations.\cite{moessner_theory_2003}

The transition occurs once the Boltzmann weight $e^{-\beta \epsilon_\alpha}$ of placing a dimer on one edge
exceeds the sum of the weights on the remaining two edges, e.g., $a \ge b + c$.
Here, $\beta$ is the inverse temperature, $\beta = 1/k_B T$, with the Boltzmann constant $k_B$ set to unity. 
$\epsilon_\alpha =-\sum_{i\in \alpha} s_i \bm{\hat e_i} \cdot \bm E_\perp$
is the energy of the local configurations of spins on a vertex $\alpha$.
The critical weight can be calculated by considering the corresponding spin configuration for each dimer
and the direction of the tilt-component of the field.

For the tilt field under consideration, $\bm E_\perp \parallel [100]$, the critical condition is
\begin{equation}
	a_c = 1 + \frac{1}{a_c} \implies a_c = \frac{1}{2} \left( 1 + \sqrt{5} \right) \approx 1.618~. \label{eq:critfug_a}
\end{equation}
In our simulation, we normalize the dipole moment of the spin $\bm d_i = s_i \bm{\hat e_i}$.
In doing so, the critical temperature $T_K$ reads as
\begin{equation}
    T_K = \frac{-\sum_i s_i \bm{\hat e_i} \cdot \bm E_\perp}{\log a_c} = \frac{\epsilon_a}{\log a_c} = 3.396~, \label{eq:TK_Et01}
\end{equation}
where $\epsilon_a$ is the energy of the spin configuration that is equivalent to the energy of a dimer at the edge with weight $a$. 

\subsection{Intermediate plateau for $\bm E \parallel [010]$: Dimer model on a square lattice} \label{scn:ice_sq_dim}

\begin{figure}[tb]
    \includegraphics[]{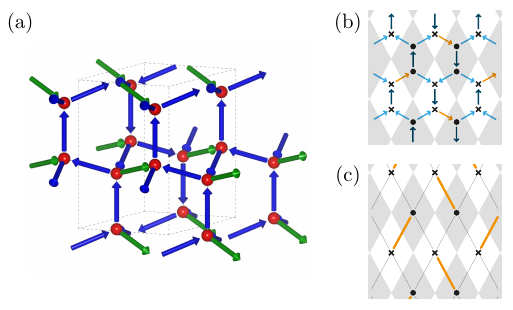}
    \caption{
        Upon applying a field along $[010]$ and lowering the temperature below the corresponding energy scale,
        the spins with the largest projection onto the field get pinned.
        These are highlighted as green spins in (a).
        The remaining degrees of freedom, blue spins in (a),
        form a non-bipartite honeycomb-like lattice
        with two neighboring vertices being of the same two-in-one-out configuration (or vice versa)
        as illustrated in (b).
        Upon contracting such a double vertex,
        a one-to-one map to dimers on the edges of a rhombic lattice (c) exists.
    }
    \label{fig:ice_spins010}
\end{figure}

We now turn to the field direction that has no analog in pyrochlore spin ice.
If the field is applied along $[010]$,
the intermediate polarization plateau can be understood
as layers of hardcore dimers on a rhombic lattice.
Each layer is separated by the spin sublattice that is pinned by the field [green spins in Fig.~\ref{fig:ice_spins010}(a)].
Thermal fluctuations will occur only within two-dimensional layers.

Each layer is composed of double vertices with two neighboring nodes of the same kind,
either both of three-in-one-out or both of one-in-three-out configuration.
See also Fig.~\ref{fig:ice_spins010}(c).
By combining two such nodes to a double vertex one internal spin and four external spins are obtained.
Three of the external spins are pointing out (in) and one in (out).
Thence, a one-to-one mapping to dimers exists by assigning a dimer to each minority spin.
In doing so, one ends up with hardcore dimers on a rhombic lattice, see Fig.~\ref{fig:ice_spins010}(c). 

Such a dimer model still exhibits an exponentially growing
ground state degeneracy upon increasing system size.
The residual entropy is reduced, but remains finite
and acquires a value of $S/k_B = G/\pi \approx 0.291$ per site,
where $G$ is Catalan's constant.\cite{fisher_statistical_1961,kasteleyn_statistics_1961}

A tilted field removes the remaining entropy.
Similar to the hexagonal case discussed before, 
a concrete condition for the ordering transition on the square lattice is known:\cite{cohn_variational_2001}
the dimers order once the Boltzmann weight $e^{-\beta \epsilon_i}$ of a dimer 
exceeds the sum of the weights of the other three dimer orientations.
Note that due to the field, the dimers with the same orientation, horizontal or vertical, come in pairs with inverse weight.
By swapping all spins of a given configuration its energy changes sign resulting in an inverse weight. 

Assuming the ordering of weights $a > b \ge \frac{1}{b} > \frac{1}{a}$,
the critical condition reads $a = b + \frac{1}{b} + \frac{1}{a}$.
By inserting the corresponding energies, cf. Eq.~(\ref{eqn:energy_doublevertex}),
we arrive at a transcendental equation
\begin{equation}
	\sinh\left(-\frac{\epsilon_a}{k_B T_K}\right) = \cosh\left(-\frac{\epsilon_b}{k_B T_K}\right)~.
\end{equation} 
As before, $k_B$ is set to one.
For a tilt field $\bm E_\perp$ applied along $[101]$, we obtain a critical temperature of
\begin{equation}
    \frac{T_K}{E_\perp} \approx 3.385~\label{eq:TK_Et1t} 
\end{equation}
in units of the tilt field $\bm E_\perp$.
$T_K$ is shown in the polarization plot, cf. Fig. \ref{fig:ice_P_C_sq},
as a dashed vertical line and agrees with the simulation.

In the following section, the Kasteleyn theory is applied to obtain exact expressions results for the effective dimer model. 

\section{Kasteleyn-Matrix Approach for $\bm E \parallel [010]$: Dimers on a Square Lattice} \label{scn:ice_dimer}

\begin{figure}[tb]
    \includegraphics[]{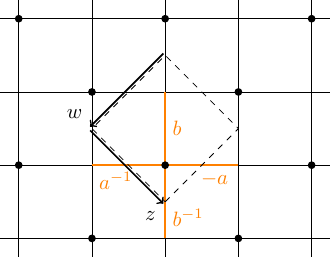}
    \caption{Unit cell convention for the computation of the partition sum
        utilizing a Kasteleyn matrix with bond weights and corresponding bonds highlighted in orange.}
    \label{fig:kasteleyn_sq_dim_convention}
\end{figure}

The general goal is to provide an expression for the partition function of hardcore dimers densely packed on edges of a graph~\cite{kasteleyn_dimer_1963}
\begin{equation}
    \mathcal Z = \sum_{\{n_1 ,n_2 ,\ldots,n_N \}} g(n_1, n_2, \cdots,n_N ) z_1^{n_1} z_2^{n_2} \cdots z_N^{n_N} ~,
\end{equation}
where the Boltzmann weight $z_i=e^{-\beta E_i}$ belongs to edge $i$ with an occupation number $n_i= \{0,1\}$,
and $g(~)$ is a generating function.

The partition sum $\mathcal Z$ for dimers on any planar graph can be written in terms of a Pfaffian~\cite{kasteleyn_dimer_1963} 
\begin{equation}
    \mathcal Z = |\mathrm{Pf}~K|
    \label{eqn:partsum_pfaffian}
\end{equation}
of a skew-symmetric matrix $K$. 
The Pfaffian of a skew-symmetric matrix is defined via the determinant as $(\mathrm{Pf}~A)^2 = \det A$.
The matrix $K$ is the directed adjacency matrix of a graph---here lattices---%
with weights $z_i$ and an additional phase factor $\phi_i$.
$\phi_i$ has to be chosen such that a loop around a face with $0 \mod 4$ edges acquires a negative sign
and a positive sign around a face with $2 \mod 4$ edges,\cite{kasteleyn_dimer_1963}
This ensures, that all the terms generated in $\mathrm{Pf}~K$ have the same sign
and Eq.~(\ref{eqn:partsum_pfaffian}) is valid.  

A simplification exists for bipartite lattices:
If the nodes of the lattice are numbered such that the nodes of sublattice~A get indices $\{0,1,\ldots,N/2-1\}$
and the nodes of B indices $\{N/2,\ldots,N-1\}$, then $K$ has the structure
\begin{equation}
    K = \begin{pmatrix}
            0 & \tilde K \\
            -\tilde K^T & 0 \\
        \end{pmatrix}~.
\end{equation}
Here, $\tilde K$ denotes the Kasteleyn matrix, which is the
(reduced) adjacency matrix.
The matrix $\tilde K$ incorporates only the connections from even to odd sites, 
i.e. rows of the adjacency matrix represent nodes of sublattice~A and columns represent nodes of B.
In contrast to $K$, the reduced $\tilde K$ is not defined with respect to a directed lattice.

Since {$\det~K=(\det~-\tilde K^T)(\det~\tilde K) = (\det~\tilde K)^2$,} %
Eq.~(\ref{eqn:partsum_pfaffian}) simplifies to
\begin{equation}
    \mathcal Z = |\det~\tilde K|~.
    \label{eqn:partsum_pfaffian_bip}
\end{equation}
%
%
Figure~\ref{fig:kasteleyn_sq_dim_convention} illustrates the convention
of the fundamental unit cell and the edge weights.
With this choice, $\tilde K$ reduces to a scalar and reads
\begin{equation}
    \tilde K = b + \frac{zw}{b} - aw + \frac{z}{a} = \det \tilde K~,
\end{equation}
with complex variables $z = e^{i\phi}$ and $w = e^{i\psi}$.

Following Refs.~[\onlinecite{kenyon_dimers_2006,kenyon_lectures_2009}], 
the partition sum, $\mathcal Z$, can be expressed as a double integral in the complex plane
\begin{multline}
    \frac{1}{N}\log \mathcal Z = \\ \frac{1}{(2\pi i)^2} \oint\limits_{|w|=1} \frac{dw}{w} \oint\limits_{|z|=1} \frac{dz}{z}
                \log \left(b + \frac{zw}{b} - aw + \frac{z}{a}\right)~.
    \label{eqn:sq_dim_partsum}
\end{multline}
A direct computation of (\ref{eqn:sq_dim_partsum}) for arbitrary weights $a$ and $b$ is difficult, if feasible at all.
The logarithm, $\log z$, of a complex variable $z$ has a branch cut for $\{z \in \mathbb R, z \le 0\}$
and, thus, the residue theorem cannot be employed.

However, given the partition sum, thermodynamic quantities are defined as derivatives of $\log \mathcal Z$,
resulting in terms suitable for the residue theorem.
Let us focus on the density of dimers on a horizontal bond.
If a dimer is placed right or left of an even site (filled dot in Fig.~\ref{fig:kasteleyn_sq_dim_convention},
its weight is inverted, $a \rightarrow a^{-1}$.
Thence, the densities on either of the horizontal bond are correlated
and its difference can be derived
\begin{equation}
    \rho_a = \frac{N_a - N_{a^{-1}}}{N} = a \frac{\partial}{\partial a} \log \mathcal Z~. \nonumber
\end{equation}
By moving the derivative $\partial/\partial a$ inside---the integration paths do not depend on $a$---we get
\begin{equation}
    \rho_a = \frac{1}{(2\pi i)^2} \oint\limits_{|w|=1} \frac{dw}{w} \oint\limits_{|z|=1} \frac{dz}{z}~
             \frac{-aw - \frac{z}{a}}{b + \frac{zw}{b} - aw + \frac{z}{a}}~.
    \label{eqn:kasteleyn_sq_rho_partition_sum}
\end{equation}
The inner integral with respect to $z$ is performed using the residue theorem.
First order poles are situated at
\begin{equation}
    z_0 = 0 \qquad z_1 = ab \frac{aw-b}{aw+b}~,
\end{equation}
of which $z_0$ is always within the integration path,
and $z_1(w)$ depends on the second integration variable.
Along the integration path, $|w|=1$, of the outer integral,
$z_1(w)|_{|w|=1}$ is either completely inside, partially inside,
or completely outside of the integration path, $|z| = 1$, of the inner integration.
\begin{figure}
    \includegraphics[]{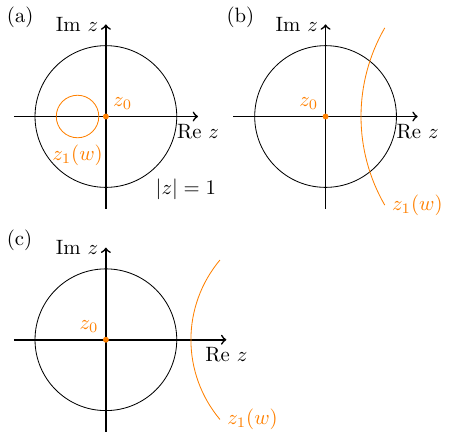}
    \caption{
        Poles of the integral in Eq.~(\ref{eqn:kasteleyn_sq_rho_partition_sum}) with respect to $z$.
        Three different cases exist depending on the choice of weights $a,b$:
        (a) $z_1(w)$ is completely inside,
        (b) $z_1(w)$ is partially inside,
        and (c) $z_1(w)$ is completely outside of the integration contour $|z| = 1$.
    }
    \label{fig:kasteleyn_integration_contours}
\end{figure}
See Fig.~\ref{fig:kasteleyn_integration_contours} for an illustration of the three different cases.
Which of the three cases applies depends on the choice of the weights $a,b$.
The cases of $z_1(w)$ being either completely inside or outside are trivial
in that the residue theorem can also be applied for the integration with respect to $w$.
These cases occur if one of the weights is larger than the three remaining ones, e.g., $a > b + \frac{1}{b} + \frac{1}{a}$,
and the dimer configuration is long-range ordered.
Consequently, the density $\rho_a$ is constant
\begin{equation}
    \rho_a = 
    \begin{cases}
        ~ ~1 \quad &\text{if} \quad (a > b + \frac{1}{b} + \frac{1}{a}) \\
        ~ ~0 \quad &\text{if} \quad (b > \frac{1}{b} + a + \frac{1}{a})~\text{or}~(\frac{1}{b} > b + a + \frac{1}{a})\\
        ~ -1 \quad &\text{if} \quad (\frac{1}{a} > b + \frac{1}{b} + a)~. 
    \end{cases}
\end{equation}

In the remaining case the pole $z_1(w)$ is partially inside $|z|=1$.
Thence, the outer integral does only run along the path of $w$ for which $|z_1(w)|<1$ resulting in
\begin{multline}
    \rho_a =  \frac{1}{2\pi i} \oint\limits_{|w|=1} \frac{dw}{w} ~ \frac{aw}{b-aw} \\
            + \frac{1}{2\pi i} \int\limits^{w_2}_{\substack{w_1\\|w|=1}} \frac{dw}{w} ~ \frac{-aw}{aw-b} - \frac{b}{aw+b}~.
\end{multline}
The first term comes from the residue at $z_0 = 0$, and the second term from the residue at $z_1(w)$.
The first term is easily solved using the residue theorem again: a pole $w_0=b/a$ is either inside or outside $|w|=1$
and results in a step function $\Theta(x)$, which is zero if $x \le 0$ and one if $x>0$.
For the second term, an expression of the indefinite integral exists.
Consequently, the result of the definite integral is simply the difference
of the indefinite integral at the final, $w_f=e^{i\phi_0}$,
and the initial point, $w_i=e^{-i\phi_0}$, of the path.
Thence,
\begin{equation}
    \rho_a =  \Theta\left(1-\frac{b}{a}\right)
            - \frac{1}{\pi} \phi_0 
            + \frac{1}{2\pi i} \log \frac{b^2 - a^2 + 2i ab \sin \phi_0}{b^2 - a^2 - 2i ab \sin \phi_0}
\end{equation}
with
\begin{equation}
    \phi_0 = \arccos\left(\frac{\left(a^2+b^2\right) \left(a^2 b^2 -1 \right)}{2 a b \left(a^2 b^2+1\right)}\right)~,
\end{equation}
being the angle at which $|z_1(e^{\pm i \phi})|=1$.

By mapping back to the spin model, the polarization is obtained as
$\bm P = d \sum_i \left< s_i \right> \bm{\hat e_i}$,
where $\left<s_i\right>$ is the thermal average of the Ising variable $s_i$,
which depends on $\rho_a(a,b)$ and $\rho_b(a,b)$.
Figs.~\ref{fig:ice_P_C_hex} and \ref{fig:ice_P_C_sq} include both the numerical results
and the exact values for the polarisation in the thermodynamic limit.
The latter exhibits a sharp cusp at the transition temperature.
This cusp rounds off within the numerical simulations with a more pronounced rounding effect for small systems.

\begin{figure}
    \includegraphics[width=\linewidth]{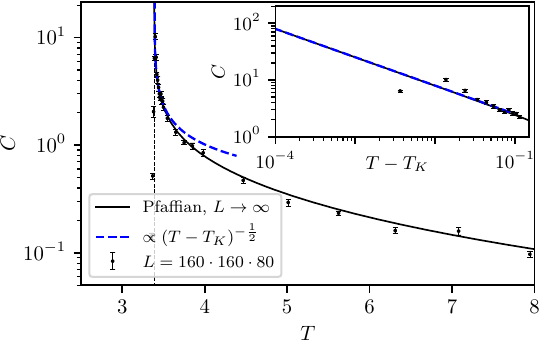}
    \caption{
        Specific heat at the Kasteleyn transition as obtained using
        (a) the Kasteleyn-matrix approach (solid black line)
        of the emergent dimer model on a square lattice
        and (b) within directed-loop Monte Carlo (black dots).
        The critical behaviour is $C \propto (T-T_K)^{-\alpha}$ with $\alpha = 1/2$ for $T>T_K$,
        as illustrated by the dashed blue line in the main plot and the inset.
        A system of size $160\cdot160\cdot80$ exhibits a maximum in the specific heat at a temperature
        slightly lifted by about $10^{-2}$ with respect to the critical temperature in the thermodynamic limit.
        Above this temperature, the numerical and analytical results agree.
    }
    \label{fig:kasteleyn_exact_C}
\end{figure}

We now turn to a discussion of the specific heat, $C$,
which is obtained in two ways:
firstly from the Kasteleyn-matrix approach by $C = \partial E/\partial T$,
where $E$ is the energy due to the Stark-coupling to the external field,
and secondly from the fluctuations of the energy within the directed-loop Monte Carlo simulation.
For $T>T_K$, $C$ follows a critical behaviour as $C \propto (T-T_K)^{-1/2}$
for a range of $T-T_K<0.1$, cf. Fig.~\ref{fig:kasteleyn_exact_C}.
The critical exponent, $\alpha = 1/2$, is in agreement with the literature.\cite{kasteleyn_dimer_1963,bhattacharjee_critical_1983}

For the finite system studied numerically, the transition shifts slightly.
Beyond the transition of the finite system, the numerically obtained specific heat has the same critical bahaviour
as extracted from the Kasteleyn-matrix approach.
For $T<T_K$ the specific heat drops to zero if being in the thermodynamic limit.
In finite systems, however, fluctuations exist near $T_K$ following an Arrhenius-type activation energy.
The energy is given by the total cost of shifting subsequent dimers along a string spanning the entire system 
and, thus, are suppressed exponentially in system size and inverse temperature.

In conclusion, the specific heat exhibits a strongly asymmetric behaviour typical for a Kasteleyn transition:
the specific heat diverges as $(T-T_K)^{-1/2}$ for $T>T_K$,
whereas $C=0$ at $T<T_K$ since fluctuations are fully suppressed
due to the hard-dimer constraint that in our case originates from the ice rules.
Finite size systems mitigate this asymmetry slightly as the energy cost to introduce defects decreases for smaller systems.

\section{Correlations and Static Structure Factor} \label{scn:ssf}

\begin{figure*}[tb]
	\includegraphics[]{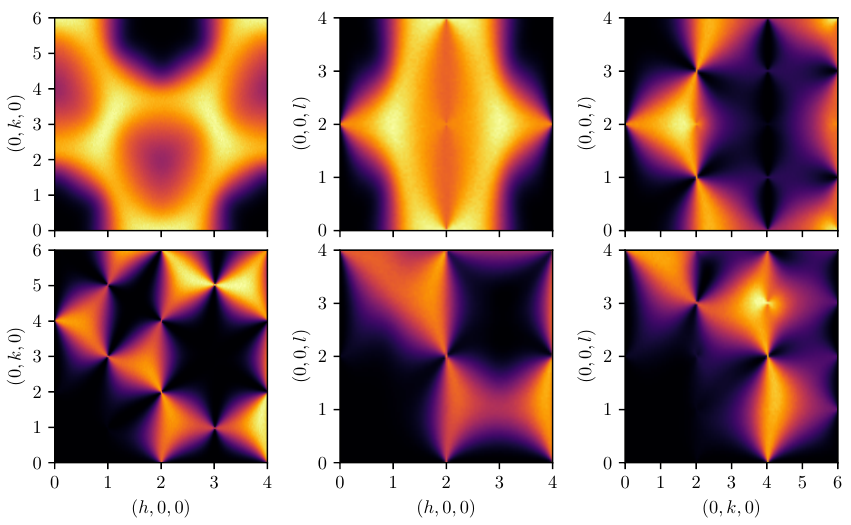}
    \caption{
        Diffuse part of the static structure factor at high temperatures
        or equivalently without external field.
        The upper row shows the $[hk0]$-, $[h0l]$-, and the $[0kl]$--plane for $S_I(\bm k)$,
        see Eq.~(\ref{eqn:ice_Sk_I}),
        of the effective spin model with Ising variables $s_i$.
        The lower row shows the same scattering planes for the proton system,
        $S_H(\bm k)$ given in Eq.~(\ref{eqn:ice_Sk_H}),
        which can be compared to experimental scattering data of ice I$_h$.\cite{li_diffuse_1994}
        The local ice rule constraint results in pinch points with typical bow tie features.
    }
    \label{fig:ice_sf_highT}
\end{figure*}

In this section, we study the static structure factor (SSF),
which allows for comparison with scattering experiments like inelastic neutron scattering.
The SSF is the Fourier transform of the equal-time two-point correlations
and reveals details about the microscopic degrees of freedom and how they are correlated.
Here, we focus on the diffuse part of the SSF capturing the contribution from the proton disorder.
The static oxygen lattice would only contribute to the Bragg peaks and is neglected.

Following the notation of Ref.~[\onlinecite{li_diffuse_1994}], the general structure factor of the protons is
\begin{equation}
 		S_{H}(\bm k) \propto \left| \sum_{j \in \text{H sites}} n_j e^{i \bm k \cdot \bm R_j} \right|^2~,
        \label{eqn:ice_SF_1}
\end{equation}
where, the sum runs over all proton sites, $\bm R_j$, of which two per link exist.
$n_j$ denotes the corresponding occupation number, which can either be zero or one.

By making use of the first ice rule, i.e. one proton per link,
the proton configuration can be rewritten in terms of Ising variables $s_j$ residing at the middle of a link, $\bm r_j$,
and the proton shift distance combined with the direction of the link, $\bm a_{\gamma(j)}$.
Here, $\gamma(j)$ maps the general label $j$ of a link to one of the distinct links with label $\beta$ within the unit-cell.
Hence, $\bm a_{\gamma(j)}$ is a purely geometrical factor,
whereas the Ising variables $s_j$ contains all the information about the degrees of freedom and their correlations.
Then, Eq. (\ref{eqn:ice_SF_1}) becomes
\begin{equation}
 		S_{H}(\bm k) \propto \left|  \sum_{j} s_j \sin \left( \bm a_{\gamma(j)} \cdot \bm k \right) e^{ i \bm k \cdot \bm r_j } \right|^2~,
        \label{eqn:ice_Sk_H}
\end{equation}
where the sum runs over all links $j$.
The sine factor leads to a suppression of any diffuse scattering within the first and partly the second Brillouin zone.

Equation~(\ref{eqn:ice_Sk_H}) simplifies when considering only effective Ising spins $s_j$
and neglecting the geometrical factor $\bm a_{\gamma(j)}$
\begin{equation}
		S_{I}(\bm k)\propto \left|  \sum_{j} s_j e^{ i \bm k \cdot \bm r_j } \right|^2~.
        \label{eqn:ice_Sk_I}
\end{equation}
We fix the convention for the $s_j$ such that a positive $s_j$
refers to a spin pointing from a site on the sublattice A to a site on B.
Even though the two SSF look very different,
they include the same information on the correlations and contain similar scattering features.

In the following, we discuss the SSF in the different regimes.
The discussion is separated into the high-temperature regime
and the intermediate regimes depending on the field direction.

\subsection{SSF at high temperatures}
The SSF at high temperatures or with small external field, respectively, is shown in Fig.~\ref{fig:ice_sf_highT},
where thermal fluctuations exceed the Stark energy and the SSF is comparable to the zero-field case. 
Good agreement is found with experimental data
based on neutron scattering~\cite{li_diffuse_1994}
and the theoretical analysis done by Wehinger et al.~\cite{wehinger_diffuse_2014},
Isakov et al.~\cite{isakov_analytical_2015}, and Benton et al.~\cite{benton_classical_2016}. 

Both SSF---that of the protons, $S_H(\bm k)$, as well as that of the spins, $S_I(\bm k)$---%
exhibit pinch points with a bow tie structure of the intensity around lattice Bragg points,
e.g. at $\bm k = (2,~2,~0)$ or $(0,~4,~2)$ for $S_H(\bm k)$ and at $\bm k = (0,~2,~1)$ or $(0,~0,~2)$ for $S_I(\bm k)$.
The bow tie structure is typical for local constraints like the ice rules,
that are equivalent to a divergence free condition on a lattice.
The ice rules translate to Gauss law, {$\nabla \cdot \bm E = 0$,}
that eliminates the longitudinal component of the field.
Regarding the microscopic degrees,
the ice rules project out any longitudinal correlation of the otherwise unconstrained proton displacements.
%
Broad features occur, that are most noticeable in $S_I(\bm k)$ within the $[hk0]$-plane,
e.g. near $\bm k = (1,~0,~1)$.
These broad features are the diffuse remnants of pinch points in the $[hk1]$-plane
and become pinch points in the effective dimer model of the intermediate regime,
due to the dimensional reduction to two dimensions.

\subsection{SSF Within Intermediate Plateau for $\bm E \parallel [001]$}

\begin{figure}[tb]
    \centering
    \includegraphics[width=\linewidth]{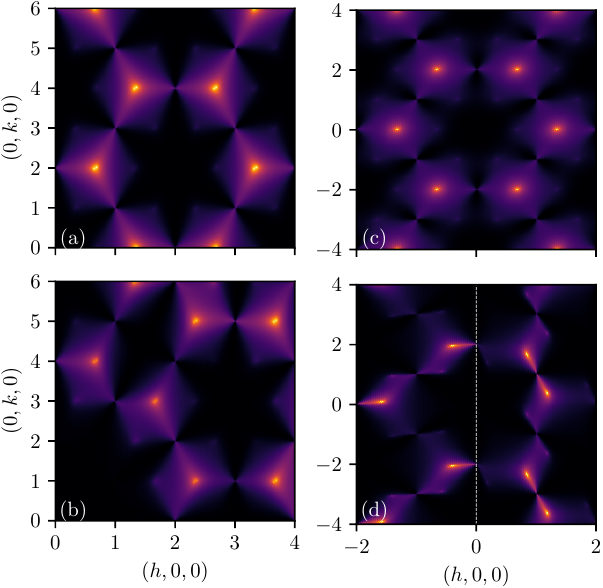}
	\caption{
        $[hk0]$ plane of the static structure factor 
        within the intermediate plateau ($T=40$)
        when a field is applied along $\bm E=(1,0,100)$ for:
        (a) effective spin model,
        (b) protons,
        (c) asymptotic correlations for dimers with equal weights on the hexagonal lattice,\cite{moessner_theory_2003}
        and (d) effective spin model at $T=5$ within the plane 
        formed by either the lower (left) or upper (right) spins of a unit cell.
        The splitting of the high-intensity points in (a) and (b)
        is caused by the small tilt component of the applied field.
        Lowering the temperature enhances the splitting as illustrated in (d)
        until two adjacent peaks merge at the concentric pinch point. 
        The merging occurs at the Kasteleyn transition.
    }
	\label{fig:sfE001}
\end{figure}
 
In the intermediate plateau upon applying a field along the $[001]$ axis,
the SSF, exhibits the sixfold symmetry of the honeycomb lattice the remaining degrees of freedom are confined to,
cf. Fig.~\ref{fig:sfE001}.
Pinch points with bow ties remain in the presence of the field.
The pinch points occur due to the local constraints of dimers, which itself is a consequence of the initially imposed ice rules.

More strikingly, the SSF exhibits peaks at $\bm k=(4/3,~0,~0)$ (in $S_I(\bm k)$) 
and equivalent points generated by the $C_6$ symmetry.
These peaks are caused by algebraically decaying correlations, $C(r) \propto \frac{1}{r^\alpha}$,
with $\alpha=2$ for classical dimers on the honeycomb lattice.\cite{moessner_theory_2003}
As such the intensity of these peaks grows logarithmically with system size.
The correlations given in [\onlinecite{moessner_theory_2003}] lead to a similar SSF
as in case of the effective spin model,%
\footnote{Yet, the structure factor plotted in [\onlinecite{moessner_theory_2003}] differs.
The intensity at and near the peaks ($k = (4/3,~0,~0)$
and symmetry-equivalent points) is suppressed in their plot.
We obtain the same SSF,
if we change a sign in the phase factor $e^{i\bm k \bm r}$ of the Fourier transform.}
cf. Fig.~\ref{fig:sfE001} (a) and (c).

When considering the SSF of the protons, cf. Fig.~\ref{fig:sfE001}(b),
the form factor due to the proton positions shift the pinch points and peaks,
yet, the qualitative features remain similar.

Upon lowering the temperature, the effect of the tilt field becomes apparent,
cf. Fig.~\ref{fig:sfE001}(d).
The peaks split and shift towards a nearby pinch point.
Two peaks superimpose in the SSF
stemming from ABAB stacking of honeycomb layers. 
The tilt field prefers different edges in the two layers.
Consequently, one peak moves towards the $\bm k = (1,~1,~0)$ pinch point
and a second peak moves towards the $\bm k = (2,~0,~0)$ pinch point.
This is illustrated in Fig.~\ref{fig:sfE001}(d), where the SSF in the left half is obtained 
by summing only over the spin variables in the lower honeycomb layer,
whereas the right half includes only the upper honeycomb layer.
Upon transitioning into the ordered phase the peak merges at the pinch point with another peak.
The two peaks form a point-symmetric pair with respect to the same pinch point.

\subsection{SSF Within Intermediate Plateau for $\bm E \parallel [010]$}

\begin{figure}[tb]
    \centering
    \includegraphics[width=\linewidth]{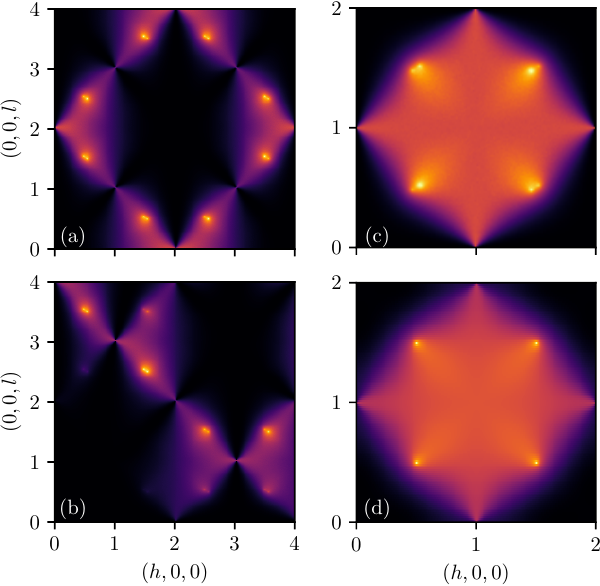}
	\caption{
        $[h0l]$ plane of the static structure factor 
        within the intermediate plateau ($T=40$) when a field is applied along $\bm E=(\sqrt 2,100,\sqrt 2)$ for:
        (a) effective spin model,
        (b) protons,
        (c) effective spin model without the intermediate spin,
        and (d) asymptotic correlations for dimers with equal weights on the square lattice.\cite{youngblood_correlations_1980}
        The splitting of the high-intensity points in (a), (b),
        and (c) is caused by the small tilt component of the applied field.
        A one-to-one map exists from the double-vertices
        with three-in-one-out (one-in-three-out) spin configurations to dimers
        enabling to compare (c) and (d).
    }
	\label{fig:sfE010}
\end{figure}

The SSF within the intermediate polarization plateau
upon applying a field along the $[010]$ axis is shown in Fig. \ref{fig:sfE010}.
The SSF of the effective spin model, $S_I(\bm k)$ as in Eq.~(\ref{eqn:ice_Sk_I}), is shown in (a), 
and the SSF of the full proton system, $S_H(\bm k)$ as in Eq.~(\ref{eqn:ice_Sk_H})), is shown in (b).

Regardless of the choice, pinch points with a bow-tie like structure occur as a signature of the ice rules.
Some of the initially broad features at high temperature sharpen and become pinch points,
e.g., at $\bm k = (3,~0,~1)$ for $S_H(\bm k)$ and at $\bm k = (1,~0,~1)$ for $S_I(\bm k)$.
Similar to the first field direction, the SSF exhibits peaks in between the pinch points,
which are caused by the algebraic decay of correlations.

In order to make the similarity to a dimer model more striking,
let us consider a SSF, where only the external spins of a double-vertex are considered.
In doing so, the reduced SSF $S_{I,red}(\bm k)$ becomes equivalent to
the SSF of dimers on a square lattice.%
\cite{fisher_statistical_1963, youngblood_correlations_1980,fradkin_bipartite_2004,tang_properties_2011}
The asymptotic correlations of dimers on a square lattice have been calculated 
for a smaller unit cell with only two Boltzmann weights,
one for horizontal and one for vertical dimers.\cite{youngblood_correlations_1980}
Nevertheless, we can use the result in the limit of equal weights which corresponds to not applying a tilt field.
Fig. \ref{fig:sfE010} allows for a comparison of the structure factor
based on our simulation (c) with the one based on the asymptotic correlations (d).
Most notably, both show the peaks at $\bm k = \left(\frac{1}{2}+n,~0,~\frac{1}{2}+m \right)$ with $n,m \in \mathbb{Z}$.
They are caused by algebraic correlation of dimers of the same orientation \cite{tang_properties_2011}
decaying algebraically with $1/r^2$.
The broad square-like feature of increased intensity has its origin in the correlation of dimers of different orientation. 
Thus, we clearly find a dimensional reduction of the three-dimensional ice 
to a tow-dimensional model of dimers on the edges of a square-like lattice.

As a remark, due to combining two oxygen nodes of different neighboring sublattices of the original ice I$_h$ lattice,
the initially defined orientation of two of the external spins is opposite to the other two spins.
This leads to a shift of $2\pi$ along the direction of alternation.
In our case this shift occurs along $[001]$ and has already been considered in Fig. \ref{fig:sfE010}(c).

\begin{table}
\centering
\caption{Location of scattering features in reciprocal space between different conventions.}

\begin{tabular}{lcc}
scattering feature~		& $\bm k$ in [\onlinecite{tang_properties_2011}] & $\bm k_{Ih,or}$ 	\\
\hline
$S_{xx}$ log. peak		& $(\pi,~0)$	& $ (\pi,~0,~\pi)$	\\
					& $(\pi,~2\pi)$	& $ (3\pi,~0,-\pi)$	\\
$S_{xx}$ pinch point 	& $ (\pi,~\pi)$	& $ (0,~0,~2\pi)$	
\end{tabular} \\
\label{tbl:id_scattering_features}
\end{table}

A second difference becomes apparent,
when considering the enlarged unit cell occurring here.
In an earlier work, only two different Boltzmann weights were considered:
one weight for horizontal and one weight for vertical dimers.\cite{youngblood_correlations_1980}
In ice I$_h$ the bipartite property of the lattice is reflected in the dimer weights
and we have to double the unit cell consisting instead of two nodes and four inequivalent edges.
A rotation of $\pi/4$ is needed in order to convert two of the small unit cells into the bigger one.
The reciprocal lattice vectors change accordingly in length and direction.
The position of the features in reciprocal space obtained in Ref.~[\onlinecite{tang_properties_2011}]
with the positions in the reciprocal space of ice I$_h$, cf. Table~\ref{tbl:id_scattering_features}.
The diffuse scattering is not affected apart from the shift and rotation.

Upon applying the tilt field, the logarithmically divergent peaks split into a pair of peaks
due to the doubled unit cell and the two species of horizontal and vertical dimers.
Furthermore, the weight of the peaks becomes asymmetric since one of the two dimers of each orientation is energetically preferred.
Its occupation number and the intensity of the corresponding peak increases.
Upon further decreasing the temperature
and approaching the Kasteleyn transition, two peaks merge at a pinch point. 
This behaviour is not captured by the asymptotic correlations derived in Ref.~[\onlinecite{youngblood_correlations_1980}].

By further reducing the temperature below the Kasteleyn temperature
the system exhibits a transition into a single ordered configuration,
which does not exhibit any diffuse scattering.

\section{Conclusion} \label{scn:con}

We studied a minimal model of the proton subsystem of hexagonal water ice I$_h$ coupled to an electric field.
The field is applied along two different easy axes,
resulting in qualitative different microscopic descriptions in terms of emergent dimers.
Upon slightly tilting the field,
the proton configuration enters a long-range ordered phase with zero entropy.
The transition into the ordered state is of Kasteleyn-type.

In case of applying the field along the $[001]$ axis,
a dimer model on the honeycomb lattice emerges.
Whereas, upon applying the field along the $[010]$ axis,
the remaining degrees of freedom are described by a dimer model on the square lattice. 
The first case is in close analogy to the intermediate magnetization plateau in spin ice with an magnetic field along $[111]$.
Its transition to the ordered state is determined in the original work of Kasteleyn.%
\cite{kasteleyn_dimer_1963}
The second case is a result of the structure of ice I$_h$
and does not have an analogy to pyrochlore spin ice.
The transition into an ordered state upon tilting the field is also 
described by the Kasteleyn matrix.
An analytic expression for the dimer densities is given,
which is directly related to polarisation, the internal energy, and the specific heat.
The polarization has a cusp at the Kasteleyn transition
and the specific heat exhibits a one-sided divergence with a critical exponent of $\alpha=1/2$ for $T>T_K$.

In both cases, the application of a field leads to a dimensional reduction of the thermally fluctuating degrees of freedom
with a simultaneously reduced power in the algebraic correlations.
In the effective two-dimensional models of dimers the correlations decay with $1/r^2$
and result in peaks in the static structure factor.
Due to the algebraic decay, these peaks diverge logarithmically with system size.

For a sharp Kasteleyn transition to occur,
strings of concurrent spins have to span the entire system along which the protons can be swapped without breaking the ice rules.
Those dynamics are unlikely to occur in a realistic ice sample, as the energy barrier is to high
and thermal activation is suppressed exponentially.
However, by introducing defects, i.e. doping with KOH,
the necessary dynamics can be restored,
but the defects will also round the characteristic cusp
of the polarization at the Kasteleyn transition.\cite{jaubert_threedimensional_2008}

Correlated quantum tunneling processes were discussed~\cite{drechsel-grau_quantum_2014,benton_classical_2016}
and may have been observed experimentally~\cite{bove_anomalous_2009,yen_dielectric_2015} in ice I$_h$ at zero field.
Within the intermediate polarization plateau, the partial ordering
restricts the quantum tunneling of protons into two-dimensional layers.
The emergent dimer models change to quantum dimer models with local ring exchange terms.
For a broad class of quantum dimer models on bipartite lattices, an ordered phase is to be expected,\cite{moessner_ising_2001}
due to the quantum order-by-disorder mechanism.
Depending on whether rotation and/or translation symmetry of the ground state is broken, e.g. staggered, columnar, or plaquette dimer configuration,
the corresponding proton configurations of each layer can be qualitatively different, e.g. ferroelectric, antiferroelectric, or paraelectric.
In contrast, the saturated plateau is completely frozen 
and quantum tunneling will be suppressed completely. 

    \section{Acknowledgements} 
    
    We are grateful to Ludovic Jaubert, and Nic Shannon for insightful discussion.
    This work was supported in part by Deutsche Forschungsgemeinschaft (DFG) via SFB 1143,
    the W\"urzburg-Dresden Cluster of Excellence on Complexity and Topology in Quantum Matter -- \textit{ct.qmat} (EXC 2147, project-id 39085490),
    and Germany’s Excellence Strategy via EXC 2111 (project-id 390814868), 
    FP acknowledges the support of the DFG Research Unit FOR 1807 through grants no. PO 1370/2- 1, TRR80,
    the Nanosystems Initiative Munich (NIM) by the German Excellence Initiative,
    and the European Research Council (ERC) under the European Unions Horizon 2020 research and innovation program (grant agreement no. 771537). 

	\appendix

	\bibliography{Ih_new.bib}
	
\end{document}